\documentclass[%
preprint,12pt
]{elsarticle}

\pdfoutput=1
\usepackage{graphicx}

\usepackage[usenames]{color}

\begin{document}
\begin{frontmatter}

\title{Surface anisotropy of iron oxide nanoparticles and slabs from first
principles : influence of coatings and ligands  as a test of the Heisenberg model}

\author{Katarzyna Brymora} 
\author{Florent Calvayrac} \ead{
Florent.Calvayrac@univ-lemans.fr} \address{
 Institut des Mol\'ecules et Mat\'eriaux du Mans, PSC IMMM, CNRS
 UMR 6283, Universit\'e Bretagne Loire
Universit\'e du Maine, Av.Messiaen, 72085 Le Mans Cedex 9, France
}

\date{\today}

\begin{abstract} We performed $ab$ $initio$ computations of the
magnetic properties of simple iron oxide clusters and slabs.  We considered an
iron oxide cluster functionalized by a molecule or glued to a gold
cluster of the same size. We also considered a magnetite slab coated
by cobalt oxide or a mixture of iron oxide and cobalt oxide.  The changes in magnetic behavior were
explored using constrained magnetic calculations.
A possible value for the surface
anisotropy was estimated from the fit of a classical Heisenberg model
on {\em ab initio}  results.  The value was found to be compatible with
estimations obtained by other means, or inferred from experimental
results. The addition of a ligand, coating, or of a metallic nanoparticle
to the systems degraded the quality of the description by the
Heisenberg Hamiltonian. Proposing a change in the anisotropies allowing
for the proportion of each transition atom we could get a much better 
description of the magnetism of  series of  hybrid cobalt and iron oxide systems.

\end{abstract}

\end{frontmatter}

\section{\label{sec:level1}Introduction}

Numerous experimental and theoretical studies have been performed  in the 
last decades to provide an understanding of the properties of
magnetic nanoparticles, such as super-paramagnetism or surface-enhanced
anisotropy\cite{kodama_magnetic_1999}\cite{lu}.

It has been proven possible to create nanoparticles built with
alternating layers of various materials exhibiting different magnetic
behaviors, which will influence the magnetic moment structure at the
surface or interface \cite{bias}. An induced magnetic anisotropy
(exchange bias) can appear in such systems, corresponding to a
shifted hysteresis loop and a ferrimagnetic alignment of the moments
near the center of the nanoparticle as well as a pinning of the
magnetic moments on the surface.

It has been demonstrated that surface modification by  organic
ligands have a strong influence on the magnetic structure of the
nanoparticles (as described in the pioneering work
of\cite{berkowitz_spin_1975}) and also cause  spin pinning near
the surface\cite{koseoglu_effect_2006}.

Surface effects alone can change the magnetic structure, as it was
demonstrated for instance in the case of cobalt
\cite{luis_enhancement_2002}, cobalt oxide \cite{hajra_enhancement_2012},
magnetite powders \cite{kihal_high_2012}, and maghemite nanoparticles
\cite{nadeem_spin-glass_2012}.  Those effects have been reviewed
in the case of iron oxide in\cite{tronc_surface-related_2000}.

From the theoretical point of view, those effects and the related
phenomena have been considered mainly from the phenomenological
side. Typically, Monte-Carlo calculations on the classical Heisenberg
model were performed\cite{van_leeuwen_quenching_1994}, and  results
were compared to experimental data obtained on CO-functionalized
NiPt clusters, as well as to DFT calculations where magnetic moments
were computed in the collinear local density functional approach.

The Monte-Carlo-Metropolis approach was also used in the case of
magnetite nanoparticles in\cite{mazo-zuluaga_surface_2009}.  In
this work, the authors have demonstrated that the surface anisotropy
constant can heavily influence the exchange-bias behavior. But this
value  remains a parameter and cannot even be precisely inferred
from experimental data; only a range of possible values is estimated
from the resulting magnetic behavior and the comparison to experiment.

The problem of the {\em ab initio}  computing of the magnetic
anisotropy at the surface of a nanoparticle will be the focus of
this paper. In literature, this parameter is a phenomenological
input in large scale classical calculations based on modified
Heisenberg models. In the present work, on the example of a small cluster, 
namely Fe$_{13}$O$_8$
we linked various magnetically constrained calculations 
to a Heisenberg model in order to
estimate magnetic properties of the nanoparticle from first principles.
We studied the change in magnetic properties due the presence of a
ligand (dopamine) or of a nearby gold cluster. We also considered
a magnetite surface, with an eventual cobalt oxide layer. We discussed the resulting
pertinence of the description by a Heisenberg Hamiltonian.
We then adjusted the same Hamiltonian on an ensemble of results obtained
on mixed cobalt oxide and iron oxide clusters.

\section{Theory/calculation}

\subsection{Structure of the Chosen Systems}

The first objective was to obtain an optimized structure of
Fe$_{13}$O$_{8}$, which according to mass spectrometry shows a higher
abundance than other iron oxide clusters with different compositions
\cite{sun2}.  We chose this system because it is small enough to be easily modelled
by repetitive first-principle calculations and yet provide interesting surface effects, and was already
structurally studied\cite{PhysRevB.59.12672}.  We used an {\em ab initio}  ultrasoft pseudopotential
scheme with a plane-wave basis as implemented in the  Quantum Espresso
(QE) suite (\cite{QE}).  The plane wave basis set was defined by an energy
cutoff of 30 Ry (408 eV), confirmed to be sufficient by the test
of convergence of the total energy.  A mixing factor of 0.17 was employed on the densities in between
each self-consistent field iteration.  Integration
in the first Brillouin zone was performed using 1x1x1 points sampling,
since the system is an isolated cluster in a large computational
cubic box of the size of 30 {\r{A}}. The GGA density functional
from PBE\cite{pbe} was used with the corresponding pseudopotentials
computed by A.dal Corso with the "rrkj3" code\cite{dalcorso} and taken from the Quantum ESPRESSO pseudopotential data base.  The optimization
procedure was conducted without any symmetry. The structural 
results are close to the one previously reported in the literature\cite{PhysRevB.59.12672}.

\begin{figure}[!h] \includegraphics[width=\columnwidth]{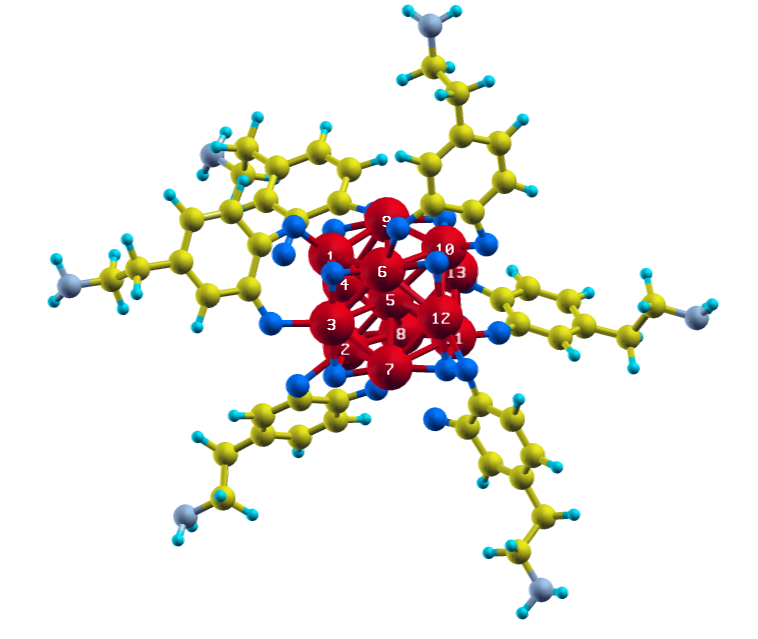}
\caption{Fe$_{13}$O$_{8}$ cluster with six dopamine molecules.}
\label{6dopa} \end{figure}

\begin{figure}[!h] \includegraphics[width=\columnwidth]{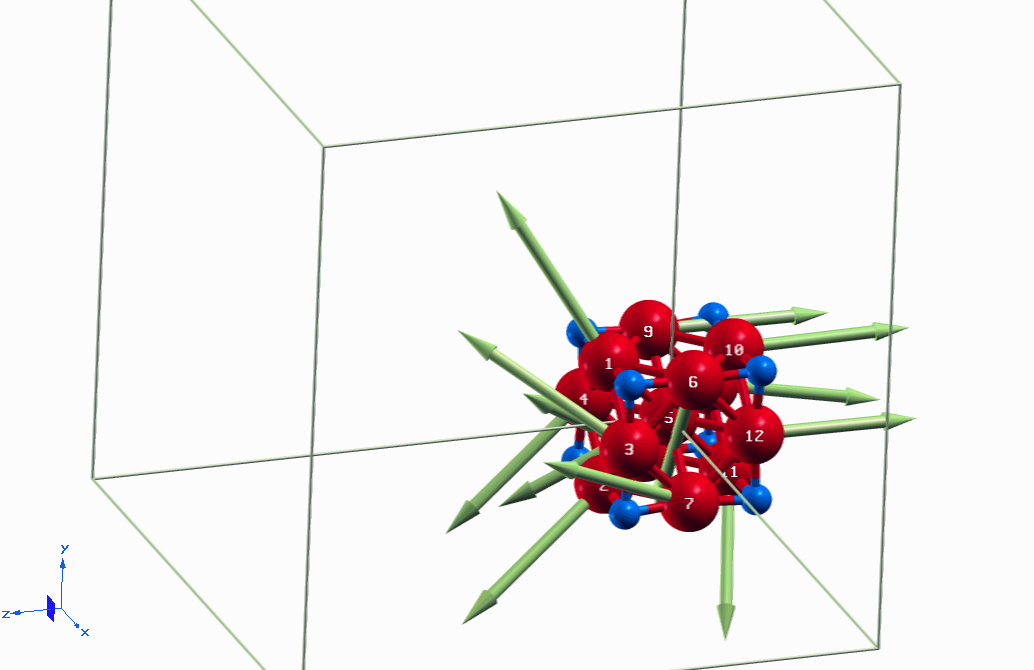}
\caption{Result of a typical non-collinear constrained calculation
of the iron oxide cluster ; here we imposed magnetic moments of 5,1
and 45 $\mu_B$ on each axis.} \label{spins} \end{figure}

\begin{figure}[!h] \includegraphics[width=\columnwidth]{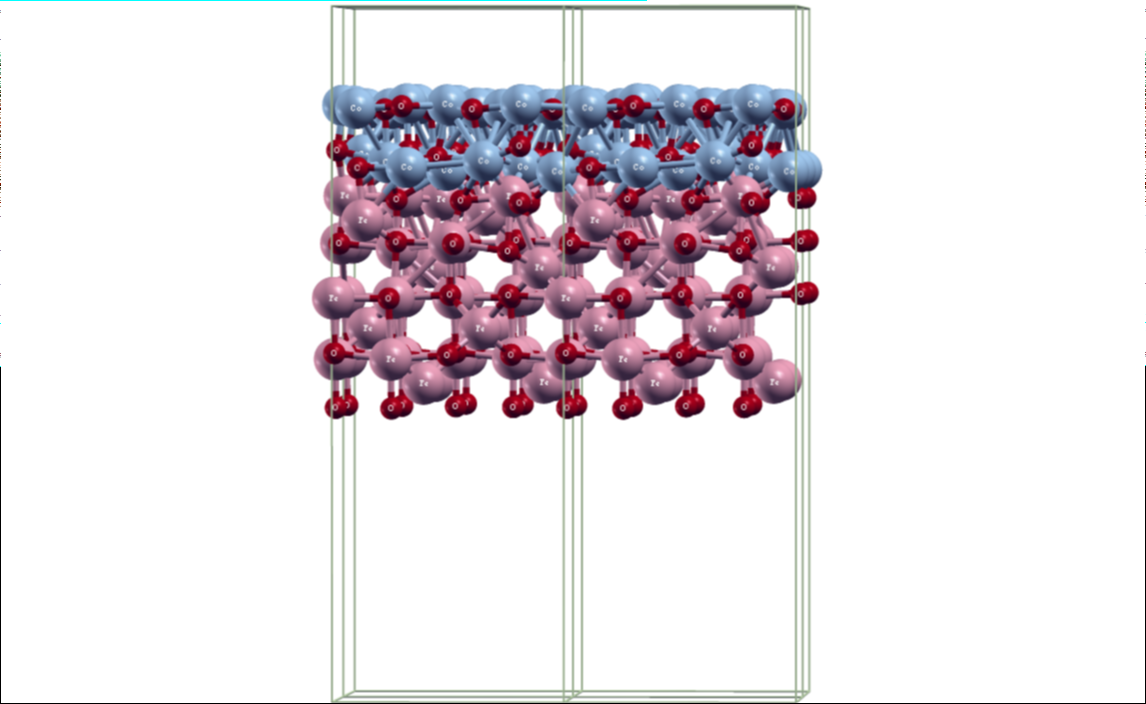}
\caption{Atomic positions in a magnetite slab coated with cobalt oxide
} \label{magneco} \end{figure}

\begin{figure}[!h] \includegraphics[width=\columnwidth]{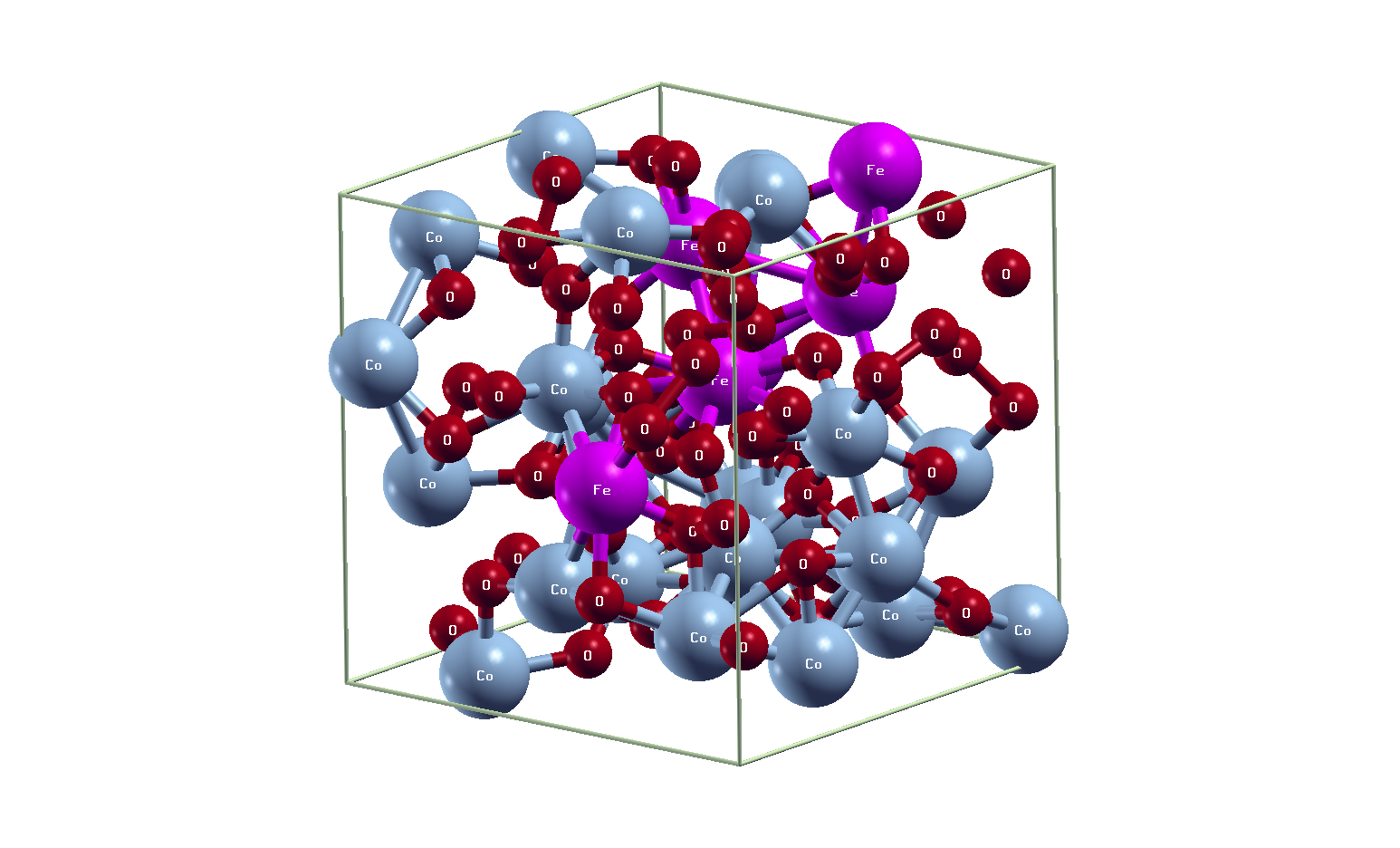}
\caption{A typical hybrid cobalt oxide and iron oxide configuration after structural optimization
} \label{cluster} \end{figure}

We also considered  two cases of infinite surfaces: 
first,  a 001 oriented slab of magnetite with
atomic positions taken from~\cite{cod}. This system 
provides an interesting approach to magnetic nanoparticles
of medical interest. The small surface to volume ratio of those
nanoparticles of a diameter close to 10 nm  makes possible the study of
the grafting of ligands  by considering a locally flat surface in order
to reduce computational cost~\cite{fouineau}. Another interest
of such surfaces lies in their catalytic properties.

In a second step,
 we coated this surface in a close
to epitaxial condition 
with 5 crystallographic units of a  012 surface of cobalt oxide,
using periodic boundary conditions, and a vacuum of 12 atomic
units on top, generated thanks to the ASE package~\cite{ase},
with the same computational conditions as previously described 
for the isolated cluster. The atomic positions, represented on
figure~\ref{magneco}, changed by less
than 2 \% in comparison to the input values after optimization, 
and the cell parameters changed by less than 1 \%. 
A potential application of such a system might be found in the pinning 
of magnetic moments, and hence a change in hysteresis curves, 
from the coupling to antiferromagnetic materials~\cite{fongang2,grecs}

From discussions with experimental teams~\cite{fouineau} developing
ferrites made of hybrid oxide nanoparticles
 we then tried a combined approach, generating a simple molecular
dynamics simulation 
of a maghemite cluster coated with cobalt oxide, 
at 1000K, based on simple, Buckhingam-like ionic potentials~\cite{fongang2} 
and extracting several representative clusters of increasing size around
a random atom. The atomic positions in these  clusters were then optimized
using the BFGS procedure within the Quantum Espresso suite with 
the same  conditions discussed above, using periodic boundary conditions 
with a cubic cell of 32 atomic units. With the largest clusters discussed
the systems are close to the bulk, the smaller ones being close to isolated systems. We think that these clusters, even
though the crystallography or the stoechiometry deviate from pure allotropic
forms, reflect what can be found in a typical molecular dynamics run. 
One should note however that before optimization of the atomic positions,
the magnetic moments around each iron or cobalt atom were found close
to zero. A typical cluster is given on figure~\ref{cluster}.

\subsection{$Ab$ $initio$ Magnetic Computation}

\indent\indent The change in local magnetic moments was computed
using the Quantum Espresso (PWSCF) code.
We would like to note that such magnetic moments are sometimes
referred to in the literature as "spins", when they are actually
expectations on the values of spin components integrated over a
sphere of reasonable but arbitrary radius centered on each atom.
For the purpose of this calculations, we used non-collinear density
functional theory\cite{barth_local_1972}.

We first considered four cases, Fe$_{13}$O$_{8}$ alone, with one dopamine
molecule added, with six dopamine molecules (Fig.~\ref{6dopa}),
and with a structurally optimized 20
atoms gold cluster in the framework of non-collinear magnetism
calculations.  We then considered a magnetite surface
coated or not by cobalt oxide, and finally, a series of hybrid iron oxide/cobalt oxide 
clusters trending towards the bulk thanks to periodic boundaries conditions.
 In the case of magnetic computations we found that
a reduction of the cutoff energy by half ensured consistency
of the results (a convergence test showed that a cutoff energy of 30
Ry is a sufficient value to compute structural properties), and a
0.17 mixing factor of densities for self-consistency was employed.  A smearing
factor of 0.02 Ry was used. We performed calculations assuming the
systems to be isolated (in the case of the cluster) with a  Martyna-Tuckerman correction \cite{martyna}.
We checked the results by performing calculations with relativistic
pseudopotentials and spin-orbit coupling which we generated from
the values suggested in the QE distribution for cutoff (1.4 and 1.6 a.u
for Oxygen with 6 active electrons and a projector on empty 3{\it
d} states), and the ultrasoft pseudopotentials without spin-orbit
as described previously. It is to be noted that the ab initio
package we used (QE) does not compute orbital  moments. 

We enforced  a set of  arbitrary magnetization states by imposing total moments different from
the ground-state result but close to it within 2 $\mu_b$ in each direction,
 and extracted the corresponding
magnetic field from the converged values, using the constrained 
magnetic calculation option from the latest version of the Quantum Espresso code,
which gives the magnetic field as a function of the distribution of magnetic moments. We used a penalty factor of 0.001 for
this purpose in order to speed up convergence under this external
constraint. When the magnetic moments
from the ground state result are used as a constraint, the magnetic field 
found is equal to zero.  

In the case of the isolated clusters, we found a magnetic structure
rather hard to describe in simple terms. In the case of the magnetite
surface, we found, as known from the literature, a slightly ferrimagnetic state
with a resulting magnetization of zero along the $x$ and $y$ axes and of 0.17 $\mu_B$ in  the
$z$ direction, orthogonal to the surface, with an absolute magnetization of 2.48 $\mu_B$.
 In the case of the magnetite surface 
coated by cobalt oxide, we found an enhancement of total magnetization, to 9.89 $\mu _B$ in the $z$
direction, as expected experimentally.
In the case of mixed iron oxide/cobalt oxide clusters, the magnetic structure
is hard to describe, depending on the chosen configuration and on the stoiechometry.
 We kept for the calculations structures where the average magnetization 
was above 0.75 $\mu _B$ per magnetic atom, the cobalt atoms inducing a reduction of total magnetization.

\subsection{Magnetic Model}

\indent\indent The previous procedure
gives a
distribution of local magnetic moments ($\vec{S}_{i}$), magnetic
fields ($\vec{H}$), and total energies ($H$). We collected
these values from various QE runs under different constraints, 
varying the total magnetization in either direction by a few $\mu_B$,
 and fitted the parameters from the
Heisenberg Hamiltonian\cite{mazo-zuluaga_surface_2009}, given as:

\begin{eqnarray} H = &-& 2\sum_{<i,j)>}{J}_{<i,j>}\vec{S}_{i}\cdot
\vec{S}_{j}\\ \nonumber &-&  K_{V}\sum_{i} \left (
S_{x,i}^2S_{y,i}^2+S_{y,i}^2S_{z,i}^2+S_{x,i}^2S_{z,i}^2 \right )\\
\nonumber &-& K_{S}\sum_{k} \left ( \vec{S_{k}}\cdot \vec{e_k}
\right )^2-g\mu_B \vec{H}\cdot \sum_{i}\vec{S_i} \end{eqnarray}

\vspace{0.4cm}

 It is to be noted here that the Zeeman energy (last term) is absent
 from total energy QE results ; the magnetic field is output separately.

The first sum involves nearest neighbors interactions between iron
atoms.  In  reference\cite{mazo-zuluaga_surface_2009}  these were
computed as from coordination numbers.  In the bulk three different
coordination numbers appear: $z _ {AA} =4$, $z_ {BB} = z_{BA} = 6$,
and $z_{AB} = 12$.  These numbers apply for the core of the
nanoparticle.  In our case, these numbers were computed from the
coordinates of the iron atoms, enforcing a cutoff radius of 3.2 \AA such that
no atom had more than 12 neighbors. This is to be compared to experimental
results obtained on cobalt ferrites for instance~\cite{carta} where
similar values are found (at room temperature compared to the present, zero
temperature study), with a maximum interatomic distance of 3.56 \AA
to ensure a coordination of 12.

The second term in  the Hamiltonian is the core cubic magneto-
crystalline anisotropy and reference\cite{mazo-zuluaga_surface_2009}
chose a value of $ K_{V} =0.002$  meV  / spin

The third term accounts for the single-ion site surface anisotropy
where the unitary vector reads

\begin{eqnarray} \vec{e_ k} =\frac{\sum _ j \vec{P_k} - \vec{P_j}}{|\sum
_ j \vec{P_k} - \vec{P_j}|} \end{eqnarray}

\vspace{0.4cm} with $\vec{P_i}$  the position vector of each iron
atom on the surface and the sum runs over iron neighbors of $j$ .

In reference\cite{mazo-zuluaga_surface_2009} the exchange parameters
were set at a value of $J_{AA} = -0.11 $ meV, $J_{BB} = +0.63$  meV,
and $J_{AB} = -2.92$  meV corresponding to a mix of ferromagnetic
and anti-ferromagnetic interactions.  Those values were taken
from\cite{uhl_first-principles_1995} where they were fitted on
$ab$ $initio$ results using a method similar in principle to the
one we presently discussed :  bulk spin waves were fitted to
non-collinear spin calculations.

\subsection{Fitting Procedure}

The results of the $ab$ $initio$ calculations
(energies) were fitted using the Monte-Carlo Metropolis result
method, based on random configurations of the parameters.

We used a  basic, proven quasi-random recurrence generator \cite{numrec}
and a  Metropolis algorithm  where we accept configuration changes
(in our case, a change of the Heisenberg parameters) $(q'_i=q_i+\Delta
q_i)$ corresponding to a $\Delta E= E'(q'_i)-E(q_i)$ change in the
virtual energy of the system.  The virtual  energy $E$ is here
a penalty function representing the distance in between the set
of energies found with $ab$ $initio$ calculations and these found with
the Heisenberg model applied to the distribution of moments.
The change  is accepted if a random number $y$ uniformly drawn in
between  0 and 1 is lower than $P(\Delta E,T)$.

To fit the $ab$ $initio$ results using the Metropolis simulated annealing
method we choose the following parameters: a reference energy $H_0$
(fit parameter), the set of  $J_{<i,j>}$ (enforcing $J_{<i,j>}=J_{<j,i>}$),
$K_{V}$, $K_{S}$ and $g$. We first used a small set of $J_{<i,j>}$
corresponding to a small cutoff in neighborhood search, then increased
the number of neighbors to increase the quality of the fit. In order
for the results to keep a physical meaning we kept the total number
of fitted parameters (maximum of 40) well under the number $N$ of samples
used to define the virtual energy or penalty function from the
energies in Rydbergs:

$$E=\frac{1}{N} \sum _ i^N | H_{\mbox{ab initio}}-H_{\mbox{Heisenberg}}|
$$.

In the case of the isolated cluster, since the set of $J_{<i,j>}$ is not
too large, we allowed those to take any value, but in the case
of the surfaces, the large number of $J_{<i,j>}$ gave an excellent fit
anytime, whatever the conditions, and the distribution of values
obtained was hard to read. Therefore, in this latter case, we used 
the same model already used in~\cite{fongang2}, namely two parameters
for each kind of super-exchange(iron-iron, iron-cobalt or cobalt-cobalt)
related to the super-exchange angle $\theta$ in between atoms separated by an
oxygen atom,  using the
relation
\begin{equation}
    J_{\theta}=J_{180^\circ}\cos^2\theta + J_{90^\circ}\sin^2\theta
\end{equation}
which can be written in the form
\begin{equation}
    J_{\theta}=J_{90^\circ} + (J_{180^\circ}-J_{90^\circ})\cos^2\theta%
    \label{eq53}
\end{equation}
where $ J_{90^\circ} $ and $ J_{180^\circ} $ are
the coupling constant corresponding
to the super-exchange angle of $ 90^\circ $ and $
180^\circ $, respectively. Those two parameters per couple of atoms (therefore six in total) are the only ones
fitted in the model with the volume and surface anisotropies.

In the case of the isolated cluster we also  modified the Heisenberg  Hamiltonian in the model
 to allow for a local change in
surface anisotropy ending up with:

\begin{eqnarray} H= && H_0 -2\sum_{<i,j>}\vec{J}_{<i,j>}\vec{S}_{i}\cdot
\vec{S}_{j} \\ \nonumber &-& K_{V}\sum_{i} \left (
S_{x,i}^2S_{y,i}^2+S_{y,i}^2S_{z,i}^2+S_{x,i}^2S_{z,i}^2 \right )
\\ \nonumber &-&\sum_{k}K_{Sk} \left ( \vec{S_{k}}\cdot \vec{e_k}
\right )^2-g\mu_B \vec{H}\cdot \sum_{i}\vec{S_i} \end{eqnarray}

We also had to chose several parameters of the simulated annealing
procedure such as the fictitious temperature, the annealing law, and
the dependence of random changes to the temperature.

All the software used is archived on 
\begin{verbatim}
http://perso.univ-lemans.fr/~fcalvay/surf_anis_code
\end{verbatim}

\section{Results and Discussion}

\subsection{Iron Oxide Clusters}

\indent\indent We generated sets of $N=101$ configurations randomly
drawn from  {\it ab initio} results corresponding to a total enforced
magnetization running from $l$ to $5$ $\mu_B$  along the $x$ axis,
$1$ to $5$ along the $y$ axis, and $37$ to $41$ $\mu_B$  along the
$z$ axis.

\begin{figure}[!h] \begin{center}
\includegraphics[width=\columnwidth]{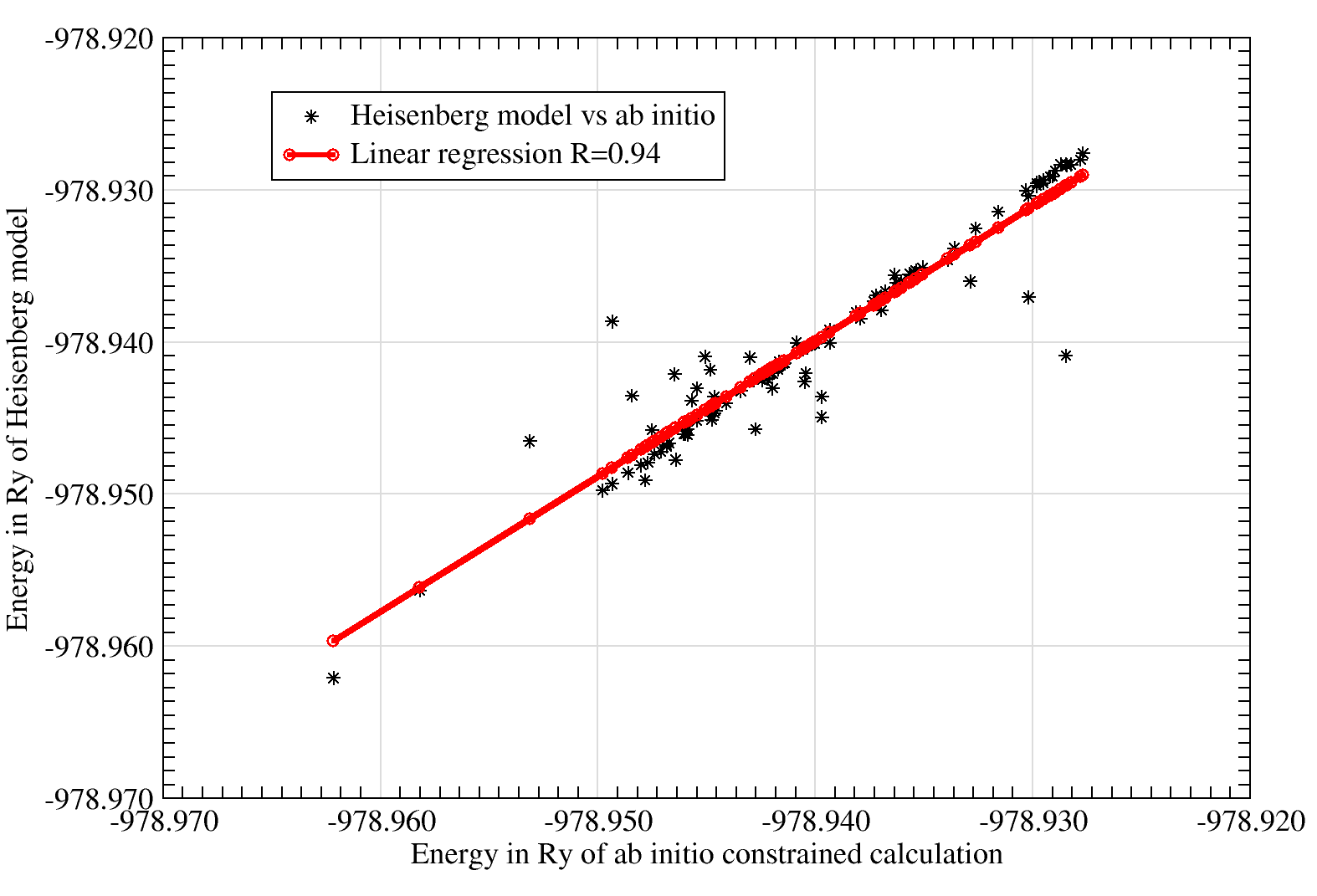} \caption{Results
of the total energy $ab$ $initio$ calculations fitted using the Monte-Carlo
Metropolis (without spin-orbit)in the case of the isolated Fe$_{13}$O$_8$ cluster.} \label{fe13o8} \end{center}
\end{figure}

The best result (penalty function as defined above less than
$10^{-4}$Ry) was, as could be expected from the increased number
of degrees of freedom in the model, 
 found using the largest number of fitting parameters
(full set of $J_{<i,j>}$ and set of surface anisotropy constants
$K_{Sk}$). A typical fit of {\it ab initio} values versus Heisenberg
model is given on Fig.~\ref{fe13o8}; in this case it can be seen
that the Heisenberg model seems to model correctly the $ab$ $initio$
results. A further confirmation is found in the fact that the $g$
factor is obtained with a value lower than  $10^{-2}$ confirming
that the absence of Zeeman energy in the QE code results is found
by the fitting procedure.

\begin{figure}[!h] \begin{center}
\includegraphics[width=\columnwidth]{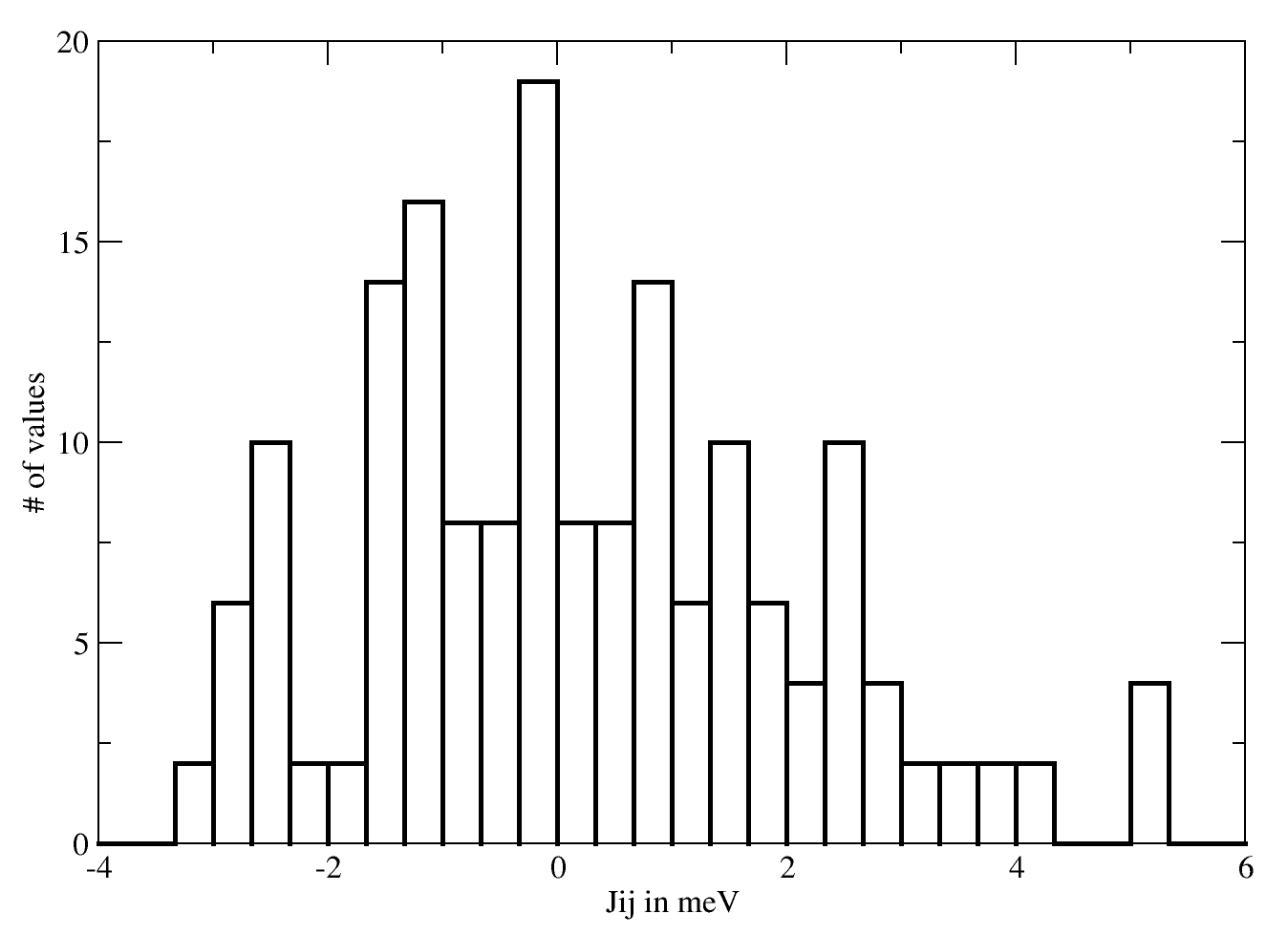} \caption{Histogram
of exchange constants found by the fitting procedure on the iron
oxide cluster.} \label{histo} \end{center} \end{figure}

A histogram of the obtained parameters for the $J_{<i,j>}$ is illustrated
on Fig.~\ref{histo}. It can be seen that those values are close
to the ones of reference\cite{uhl_first-principles_1995}, with an
alternation of ferromagnetic and anti-ferromagnetic couplings.

The volume anisotropy was found to have a value of $-2 \times 10^{-05}$
a.u, coherent with the one used in\cite{mazo-zuluaga_surface_2009}.
The values for the local surface anisotropy constants are given in
Table~\ref{surfanis}. It can be seen that the best fit corresponds
to an alternation of positive and negative values for those constants.
This fact raises the question of the validity of the Hamiltonian
used in\cite{mazo-zuluaga_surface_2009} in which a constant surface
anisotropy was used.

\begin{table}[h!] \begin{center} \begin{tabular}{|c|c|}
		     \hline
\textbf{Fe atom}                    & \textbf{Surface anisotropy
in a.u.}    \\ \hline

1 & 0.5592594$\times 10^{-01}$ \\ \hline 2  &-0.5450376$\times 10^{-01}$\\ \hline
 3 & 0.4755301$\times 10^{-01}$\\ \hline 4 & -0.2952414$\times 10^{-01}$\\ \hline 5 &
 0.4511738$\times 10^{-01}$\\ \hline 6 & 0.5525300$\times 10^{-01}$\\ \hline
7  & -0.4852317$\times 10^{-01}$\\ \hline
 8 &  -0.2241407$\times 10^{-01}$\\ \hline 9 & 0.8430323$\times 10^{-01}$\\ \hline
10 & -0.7067873$\times 10^{-02}$\\ \hline 11 & -0.7141519$\times 10^{-01}$\\ \hline 12 &
-0.7437612$\times 10^{-01}$\\ \hline 13 & -0.8089262$\times 10^{-01}$\\ \hline
			 \end{tabular} \caption{Local surface
			 anisotropy constants \label{surfanis}}
\end{center}
		 \end{table}

		 Indeed, we found that the model does not adjust
		 as well when we
use a constant surface anisotropy. The best penalty was found at a
value of $1.13	\times 10^{-3}$ Ry. This corresponds to a value of
the surface anisotropy of  1.753 a.u  and a volume anisotropy of
$6.31 	\times 10^{-05}$  a.u. The latter value is now positive, but
the authors of\cite{mazo-zuluaga_surface_2009}  found that the
$K_S/K_V $ ratio is the more important parameter to predict the
magnetic structure of a nanoparticle of intermediate size.  Here,
the value of 27780 we find  for this ratio hints at a  hedgehog
type  magnetic structure (Fig.~\ref{spins}). It might be that the large surface
contribution of such a small system as the one we study is responsible
for this fact,  but at least the value we find is not a free, almost
unknown parameter as in the literature.  Physically, such a result
would correspond to jumps during magnetization reversal, and exchange
bias properties. Such results have been experimentally observed and
are also reviewed in\cite{mazo-zuluaga_surface_2009}.

When we tried the same method with full spin-orbit coupling and
relativistic pseudopotentials, the increased disorder in the magnetic
moments resulted in a penalty function of $2 	\times 10^{-3}$ Ry,
with essentially similar results for the fitting parameters.

 \begin{figure}[!h] \begin{center}
\includegraphics[width=\columnwidth]{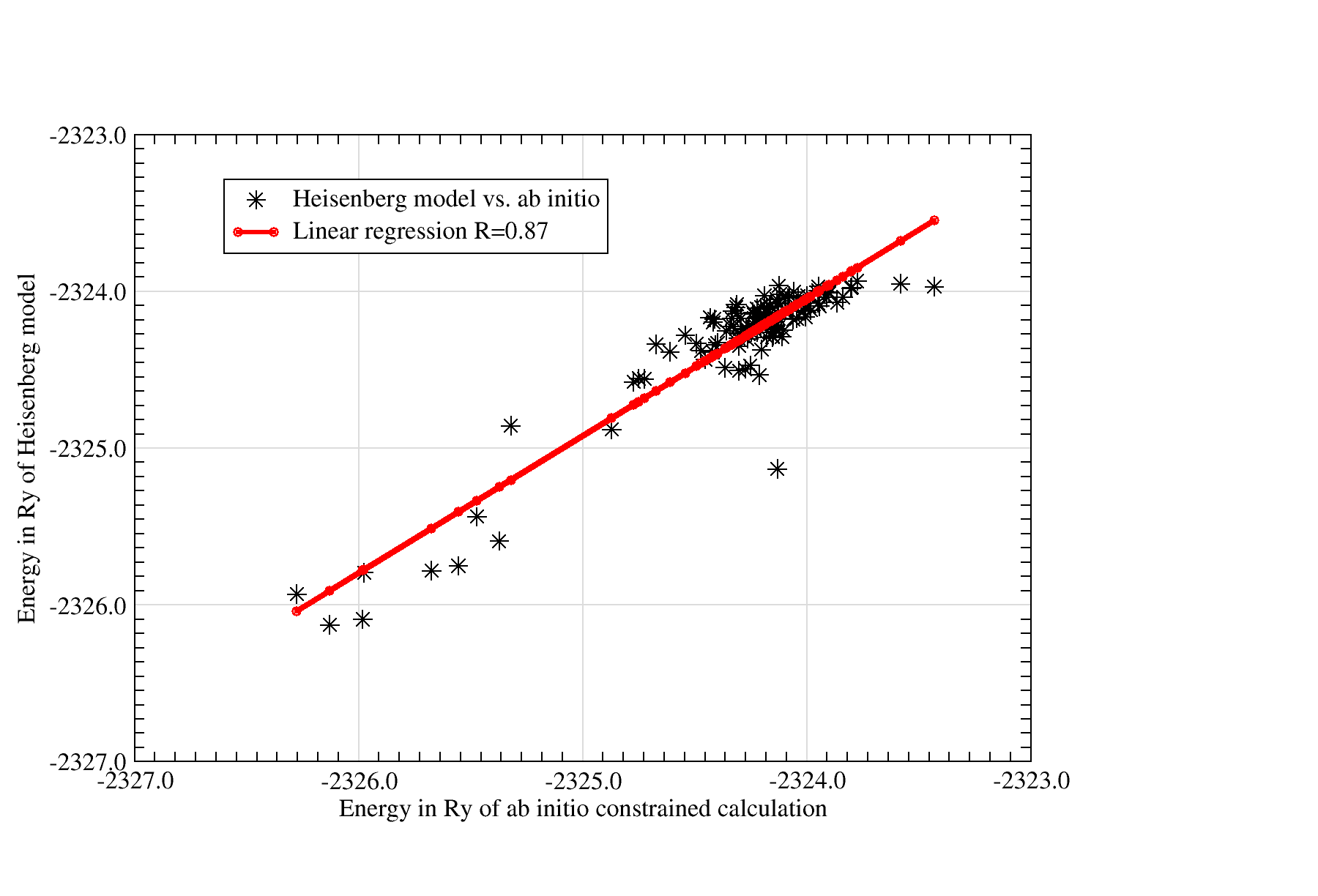} \caption{Results
of the total energy $ab$ $initio$ calculations fitted using the Monte-Carlo
Metropolis in the case of the magnetite surface} \label{magnetite} \end{center}
\end{figure}

\begin{figure}[!h] \begin{center}
\includegraphics[width=\columnwidth]{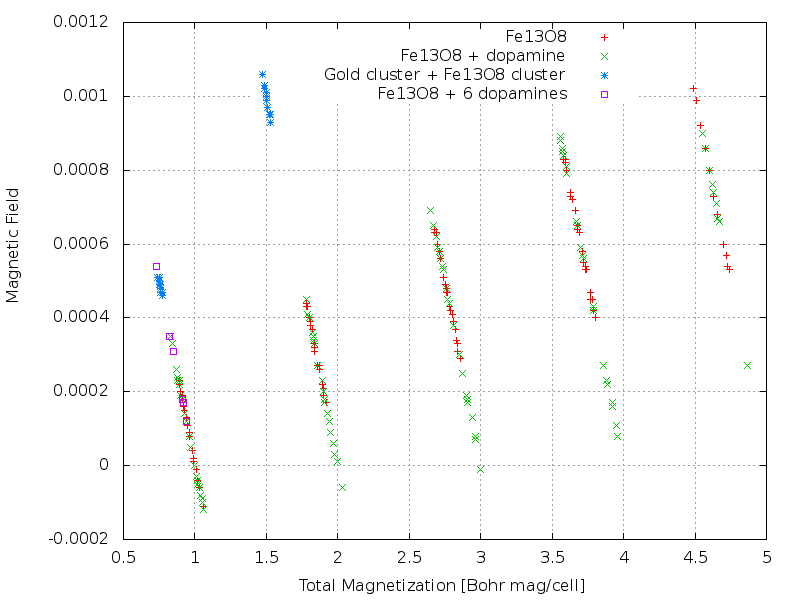} \caption{Magnetic moments in a.u.
vs. magnetic field in a.u., the latter extracted from the constrained calculation for the studied clusters
} \label{magn} \end{center}
\end{figure}

\subsection{Iron Oxide Clusters and Dopamine}

\indent\indent Linking to previous work~\cite{fouineau} on the coating of magnetic nanoparticles
by various ligands in order to make them more biocompatible or to link to antibodies in order
to better target tumors, we then added a dopamine molecule next to the cluster
from the previous section and, after optimizing the atomic positions,
tried the same fitting procedure with a Heisenberg model. It turned
out that we could not achieve a fit with a penalty function better
than $5	\times 10^{-3}$ Ry, which more or less corresponds to the
distribution of energies in the $ab$ $initio$ results, and unrealistic
coupling constants as well as a strongly unstable distribution of
surface anisotropy constants.  We concluded that a Heisenberg model
might be too simple to describe such a system, where electrons donated
by the dopamine molecule can lead to some itinerant magnetism, or
at least to some symmetry breaking.
The accuracy of $ab$ $initio$ results could also maybe be improved
to the expense of computational time, but since the fitting procedure
works in isolated cases the origin of the lack of fit seems to come
more from the model than from the quality of the results on which it
is fitted.

In order to address the question of the symmetry breaking, we added 6 dopamine molecules
(Fig.~\ref{6dopa}) symmetrically distributed around the cluster.
The simulation time was found to be too large to compute as many
constrained points as in the previous section, but, by plotting the
magnetization as a function of  the magnetic field, as can be seen in
Fig.~\ref{magn}, we could check that the response of the
system seems to be unchanged from this functionalization.


\subsection{Iron Oxide Clusters and Gold Cluster}

\indent\indent We also introduced in the system 
a small gold cluster (which could act
as a nano-antenna in plasmonics) which was found after 
structural optimization to adhere to the iron oxide cluster.
In this case, the Heisenberg model did not fit very well either
the computed $ab$ $initio$ values of energies, probably because the metallic
character of the system makes the Heisenberg model inappropriate. However, as can be
seen on Fig.~\ref{magn}, although the absolute values of external
magnetic field to achieve the same total magnetization along the
$x$ axis are strongly different from the previous cases (with
or without dopamine), the slope of the curve 
does not change by more than a few percent, meaning that the response of the system
is unchanged by the presence of a metallic gold cluster.

\subsection{Magnetite surface and magnetite coated with cobalt oxide}

In the case of the magnetite surface, generating 124 samples for 6 parameters,
we found the best fit, with a correlation coefficient of 0.87 in between
the ab initio results and the fit  for a surface anisotropy of 784.00 meV, a volume
anisotropy of 6.96x$10^-3$ meV, a Land\'e $g$ factor close to zero (which 
we see as a proof of correctness of our model), and coupling constants used
to generate the $J_{<i,j>}$ of 1.102 meV and 2.076 meV respectively. The results of
the fit are shown on figure~\ref{magnetite} and, in this case also,
we conclude that the Heisenberg model corrected with a surface anisotropy term
gives a good description of the magnetism of the system. The values we find
for volume anisotropy is in good agreement with the literature, and the ratio
of surface to volume anisotropy is in the range expected by experimentalists.

In the case of the magnetite surface coated by iron oxide, in which we could only 
generate  55 samples due to the much higher numerical cost,  we  found that the
model was much harder to fit (with a correlation coefficient of only .54  between
the ab initio values and the fitted values). The best fit was obtained 
with a surface anisotropy of 546.77 meV, volume anisotropy of 2.63x$10^{-3}$ meV, 
and coupling constants of 0.98 meV and 2.11 meV in the case of iron-iron, 0.77 meV and 2.09 meV in
the case of iron-cobalt, and 0.97 meV and 2.37 meV in the case of cobalt-cobalt pairs.
In this case, either a better ab initio
calculation would have been needed (using for instance the LDA+U method in order
to better describe the insulating behavior of cobalt oxide), or the magnetic model
has to be improved, including for instance interface anisotropy terms.

\subsection{Mixed iron oxide and cobalt oxide clusters}

We then addressed the case of clusters of iron, cobalt and oxygen atoms
of increasing size, going close to the bulk, which we submitted to a global magnetic constraint,
in this case, rather than enforcing total moments like in the previous sections
we found that the quantum calculations converged faster when
imposing  a global angle of respectively 5,30,45 and 75 degrees with respect
to the $z$ axis, cases to which we will refer as numbers 0 to 3 in the figures.
This constraint allows to get a value of the magnetic field from the ab initio calculations.
In this case we adjusted the difference in between the total ab initio energy
taking into account the spin components in the wavefunctions and the 
ab initio energy without magnetism. Here, surface atoms (for the surface anisotropy 
term) are defined by the criterion that the number of neighbors is inferior to the values
found in the bulk.

In order to estimate the change in the anisotropy term from 
the cobalt/iron mixing, after various functional tries, 
we found that the best fit was  
obtained when 
we allowed a change in the volume and surface anisotropy term depending
on the number of iron and cobalt atoms present in the cluster. The additional term
was multiplied by adjustable parameters $a_i$ (where $i$ represents
iron and cobalt atoms), so if there are $N$ different species in the system 
with $n_i$ atoms of each kind per cell for the volume term  or per surface site in the case of surfaces, 
the volume and surface anisotropies read  

\begin{equation}
 K_{S,V,i} = K^0_{S,V}a_i ^ {n_i}
\label{stoecanis}
\end{equation}

so that the final Hamiltonian we used reads 

\begin{eqnarray}
	H = &-& 2\sum_{<i,j)>}{J}_{<i,j>}\vec{S}_{i}\cdot \vec{S}_{j}\\
	\nonumber &-&  \sum_{i}  K_{V,i} \left (
S_{x,i}^2S_{y,i}^2+S_{y,i}^2S_{z,i}^2+S_{x,i}^2S_{z,i}^2 \right )\\
\nonumber &-&
\sum_{k}  K_{S,k}\left ( \vec{S_k}\cdot \vec{e_k}
\right )^2-g\mu_B \vec{H}\cdot \sum_{i}\vec{S_i} 
\end{eqnarray}

\vspace{0.4cm}
The advantage of the modification of the volume and surface anisotropies we used 
in equation~\ref{stoecanis} is that such a term allows simply to take 
into account the changes in anisotropy due to a varying ratio of cobalt to iron
atoms. 

We tried to include a factor in the Hamiltonian in order to model
the effect of the various oxygen stoechiometries of the clusters on the
energy but for the functional form we tried (multiplying each anisotropy  term 
by the ratio of magnetic to oxygen atoms) we found that the corresponding parameter was converging to zero.

The values of the parameters converged  to values presented in table~\ref{tablemix}
\begin{table}[h!] \begin{center} \begin{tabular}{|c|c|c|c|c|}
		     \hline
\textbf{System} & 
\textbf{Parameter}  & \textbf{ Value }   &
\textbf{Parameter}  & \textbf{ Value }   
  \\ \hline
Whole &
$K^0_S$  &   749.15 meV  &
$K^0_V $ &   4.43 meV   \\ \hline
Whole &
$ a_{S} $ &   2.20  &
$ a_{V} $ & 1.66  \\ \hline
$Fe$  &
$ J_{Fe,90^\circ} $ &   0.77 meV  &
$ J_{Fe,180^\circ} $ &   1.39 meV  \\ \hline
$Co$  &
$ J_{Co,90^\circ} $  & -1.25 meV &
$ J_{Co,180^\circ} $ &  -1.35 meV \\ \hline
$FeCo $  &
$ J_{FeCo,90^\circ} $ &   0.70 meV &  
$ J_{FeCo,180^\circ} $ &  1.17 meV  \\ \hline

			 \end{tabular} \caption{Parameters obtained for mixed iron oxide / cobalt oxide clusters
			  \label{tablemix}}
\end{center}
\end{table}

\begin{figure}[!h] \includegraphics[width=\columnwidth]{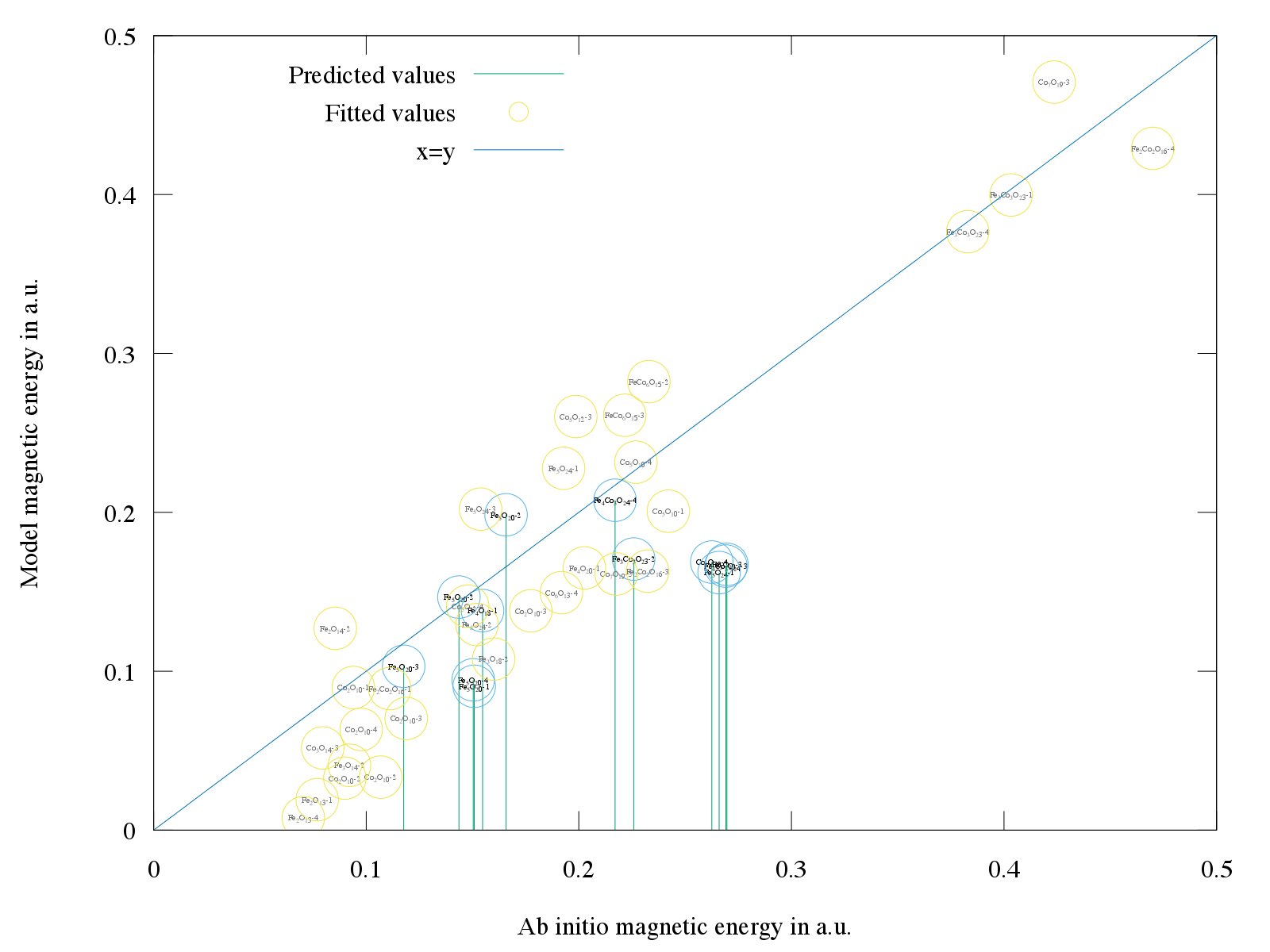}
\caption{Results of fitting the Heisenberg model on mixed iron oxide and cobalt clusters under varying magnetic constraints}
\label{configurationsfit} \end{figure}

As can be seen on figure~\ref{configurationsfit} the fitting procedure of the Hamiltonian
converged to a satisfying level, with a correlation coefficient close to 1 
in between ab initio values of the magnetic energy and energies given by the model.
We then  added, after the fitting procedure several
configurations chosen at random, in order to check the predictive capabilities of the 
model (points labeled with an impulse), which can be seen as satisfying, despite obviously less
good than the ones given by the model when reusing the input configurations used for fitting as in the previous case.  
The model can nevertheless be seen as adequate, since it uses only 12 fit parameters for 30 couples of input values,
and has some predictive power.

From the physical results one can see that the values of the exchange constants kept the same order of
magnitude as in the previous case of pure surfaces, although the values changed, with an interesting information 
for Fe/Co pairs. The same goes for the surface and volume anisotropies, with an enhancement in the case
of iron oxide in comparison to cobalt oxide, the mixed situations being intermediate.
The negative sign found in the case of cobalt oxide reflects the different magnetic order in the material,
and the values at the interface reflecting the particular magnetic properties of the interface, 
as has been studied experimentally in the case of flat surfaces~\cite{gruyters}.
One should note than in the total energies, the main contribution comes from the surface term, followed
respectively by the volume and exchange contributions. It is then no wonder that the latter term 
is particularly difficult to adjust.

Nevertheless, we can now display values of the surface anisotropy of iron oxide, and the changes induced by the 
presence of cobalt oxide.

A problem remaining to be solved is that for the time being, we do not have at our disposal a clear physical
way to predict the magnetic moments emerging from a given configuration of atoms. A machine learning
approach based on physically relevant data such as relative angles and distances can be envisaged ; 
nevertheless the present study gives surface anisotropy values from an adjustment over ab initio
results, and if values of magnetic moments are set to physically reasonable values as found
in the bulk,  can 
give a phenomenological description of magnetic nanoparticles of large size if the geometric
configurations of atoms do not diverge  too much from the bulk.

\section{Conclusions}

\indent\indent In this work, we adjusted
a classical Heisenberg model of magnetism
using the Metropolis simulated annealing method 
including surface anisotropy effects on magnetically constrained,
non collinear $ab$ $initio$ results obtained a small iron oxide
cluster functionalized or not by one or several dopamine molecules
or a nearby gold cluster. We conclude that the  Heisenberg
model seems to apply well to the simpler systems (namely, a free
iron oxide cluster or a magnetite surface), allowing us to give some absolute values of the
surface anisotropy constant, although a locally varying surface
anisotropy alternating positive and negative values seem to provide
a better description.

This could allow  us to  describe the magnetic behavior of a nanoparticle
of size  1 to 10 nm, which $ab$ $initio$ calculations  can hardly tackle
for the time being because of computing power limitations, hoping
that the large surface proportion of iron atoms in the small cluster
we have studied does not influence the results.

In the case of functionalized cluster by one or several molecules
of dopamine, or by a nearby gold cluster, the Heisenberg picture
does not apply as well as for the simpler system, but we could
nevertheless observe that the linear relation in between magnetic
field and magnetization was unchanged in all those cases even if
absolute values changed.

In the case of clusters of varying sizes and compositions, 
with periodic boundaries conditions,
we were able to give a model describing hybrid ferromagnetic
and antiferromagnetic systems, describing functionally the enhancement
in surface anisotropy due to the juxtaposition and mixing of the systems and
giving estimations of the corresponding values.

\section{Acknowledgments}

We thank S.Ammar and I.Labaye for fruitful discussions. K.B.
was supported by a grant from the French Education and Research
Minister. We thank GENCI/IDRIS and CRIHAN for computational
time (projects x2014096171 and 007 respectively)

\end{document}